\documentclass[epjST]{svjour}
\usepackage{graphics}

\begin{document}

\title{Uncertainty principle for control of ensembles
       of oscillators driven by common noise}
\author{Denis S.\ Goldobin\inst{1}\fnmsep\inst{2}\fnmsep\thanks{\email{Denis.Goldobin@gmail.com}}}
\institute{Institute of Continuous Media Mechanics, UB RAS,
             Perm 614013, Russia\and
           Department of Mathematics, University of Leicester,
             Leicester LE1 7RH, UK, EU
           }


\abstract{
We discuss control techniques for noisy self-sustained oscillators
with a focus on reliability, stability of the response to noisy
driving, and oscillation coherence understood in the sense of
constancy of oscillation frequency. For any kind of linear
feedback control---single and recursive delay feedback, linear
frequency filter, etc.---the phase diffusion constant, quantifying
coherence, and the Lyapunov exponent, quantifying reliability, can
be efficiently controlled but their ratio remains constant. Thus,
an ``uncertainty principle'' can be formulated: the loss of
reliability occurs when coherence is enhanced and, vice versa,
coherence is weakened when reliability is enhanced. Treatment of
this principle for ensembles of oscillators synchronized by common
noise or global coupling reveals a substantial difference between
the cases of slightly non-identical oscillators and identical ones
with intrinsic noise.
 }

\newcommand {\e} {\varepsilon}
\newcommand {\ph} {\varphi}
\newcommand {\lla} {\left\langle}
\newcommand {\rra} {\right\rangle}
\newcommand {\la} {\langle}
\newcommand {\ra} {\rangle}

\maketitle

\section{Introduction}
Collective phenomena in ensembles of dynamical systems can
manifest complex behavior and self-organization, which are not
possible for unitary systems with arbitrary level of complexity.
For biological systems, they extend from the simplest collective
dynamics of bacteria~\cite{bacterial_colonies} or higher
organisms~\cite{population_synchrony} to the activity of neural
tissue~\cite{Kohenen-2001,Tass-1999}.  Similarly, in technology,
essentially collective phenomena vary from plain synchronization
effects~\cite{Strogatz_etal-2005,Pikovsky-Rosenblum-Kurths-2001-2003,Zhou_etal-2002}
to the phenomena laying in the basis of operation of artificial
neural networks~\cite{Kohenen-2001,Haykin-2008}. Although these
phenomena are inherent to large ensembles, their features are
determined by characteristics which make physical sense also for
unitary systems~\cite{Pikovsky-Rosenblum-Kurths-2001-2003}.
Indeed, synchronization (and clustering) in ensembles and their
susceptibility to control are influenced by their individual
robustness properties.
``Reliability''~\cite{Mainen-Sejnowski-1995} (or
``consistency''~\cite{Uchida-Mcallister-Roy-2004}), i.e. the
stability of the system response to noisy driving, and
coherence~\cite{Pikovsky-Rosenblum-Kurths-2001-2003,Goldobin-Rosenblum-Pikovsky-2003},
i.e. the constancy of the oscillation frequency in time, are the
principal ones.

Mathematically, the natural quantifier of reliability is the
Lyapunov exponent (LE, $\lambda$) measuring the exponential decay
rate of perturbations of the system
response~\cite{Pikovsky-Rosenblum-Kurths-2001-2003,Lyapunov_exponent,Ritt-2003}.
Oscillators are more reliable for a larger negative LE.  Coherence
can be quantified by the diffusion constant (DC, $D$) of the
oscillation phase $\ph(t)$, which grows non-uniformly in time due
to either noise or chaotic dynamics;
 $\la(\ph(t)-\la\dot{\ph}\ra t)^2\ra\propto Dt$.
These quantifiers are immediately relevant for the synchronization
phenomenon in ensembles of oscillators subject to common
noise~\cite{Goldobin-Pikovsky-2005b,uncoupled_ensembles-synchronization}
or global
coupling~\cite{Strogatz_etal-2005,Topaj-Kye-Pikovsky-2001,globally_coupled_ensembles-control}.

In this paper, we report how the reliability and coherence of a
noisy limit-cycle oscillator can be efficiently controlled by a
general linear feedback.  However, the ratio of LE and DC is shown
to be independent of the noise strength and feedback parameters.
The persistence of this ratio can be formulated as an {\it
uncertainty principle}.  Finally, the important implications of
this principle for controlling collective dynamics of ensembles
are revealed---the ensemble responds to control essentially
differently for the cases of identical oscillators with intrinsic
noise and slightly non-identical oscillators with no intrinsic
noise---and conclusions are drawn. The tentative hypothesis on the
response of ensembles to the control proposed in the concluding
remarks of~\cite{Goldobin-2008} is shown to be valid only for the
case of non-identical oscillators with no intrinsic noise.

\section{General linear feedback}
Let us consider a general $N$-dimensional limit-cycle oscillator
subject to small linear delayed feedback and weak noise:
\begin{equation}
\dot{x}_i=F_i({\bf x})+a\,z_i(t)+B_i({\bf x})\circ\xi(t),
 \label{eq01}
\end{equation}
where $i=1,2,...,N$, $a$ is the feedback strength, the feedback
term
$$
z_i(t)=\int\limits_0^{+\infty}\sum_{j=1}^{N}G_{ij}(t_1)\,x_j(t-t_1)\,dt_1\,,
$$
$G_{ij}(t)$ is the Green's function, different $G_{ij}(t)$ can
feature linear ``single'' or recursive delay feedback (particular
forms of this function are specified in the text
below)~\cite{Goldobin-Rosenblum-Pikovsky-2003,Pawlik-Pikovsky-2006},
linear frequency filter~\cite{Tukhlina_etal-2008}, etc. The sign
``$\circ$'' indicates the Stratonovich form of equation; $\xi(t)$
is white Gaussian noise: $\langle\xi\rangle=0$,
 $\langle\xi(t)\xi(t')\rangle=2\delta(t-t')$;
$B_i({\bf x})$ is the susceptibility of the system to noise
driving.

\begin{figure}[!t]
\center{\sf
(a)\hspace{-4mm}\resizebox{0.31\columnwidth}{!}{%
 \includegraphics
 {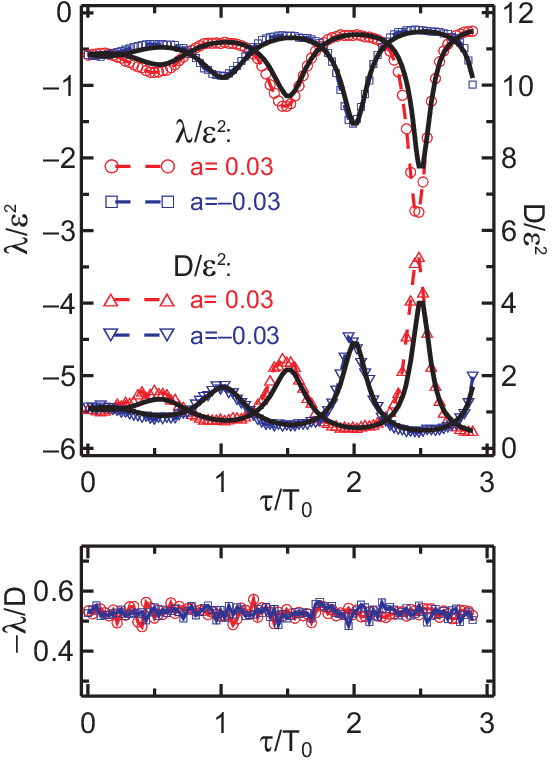} }\quad
(b)\hspace{-4mm}\resizebox{0.31\columnwidth}{!}{%
 \includegraphics
 {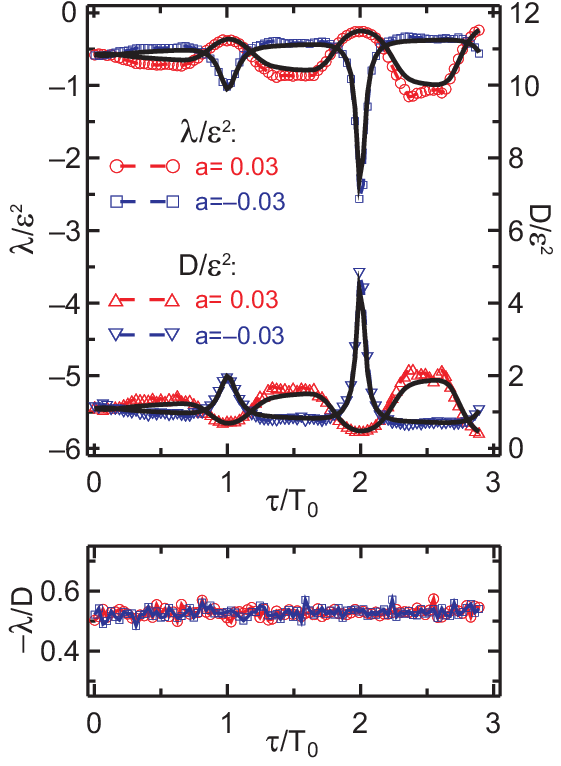} }\quad
(c)\hspace{-4mm}\resizebox{0.31\columnwidth}{!}{%
 \includegraphics
 {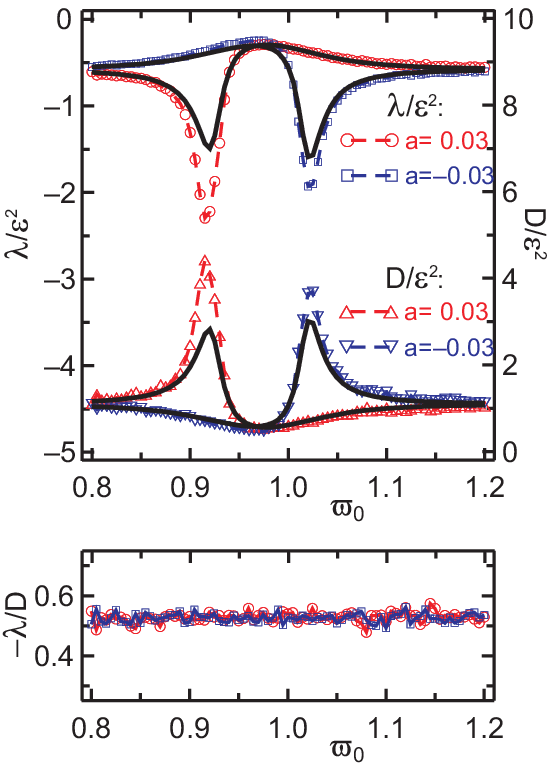} }
}
  \caption{(Color online)
Dependencies of the Lyapunov exponent $\lambda$ and the diffusion
constant $D$ (upper and lower graphs in the upper row of plots,
respectively) on the feedback/filter parameters for the van der
Pol oscillator with $\mu=0.7$ subject to Gaussian white noise of
strength $\e=0.05$. (a)~``Simple'' delay feedback
[eq.~(\ref{eq-simple})]. (b)~Recursive delay feedback
[eq.~(\ref{eq-multiple})]. (c)~Linear frequency filter
[eq.~(\ref{eq-filter})] with $\alpha=0.1$.  The oscillation period
of the control-free noiseless system $T_0\approx2\pi/0.97$\,; the
feedback strength ($a=0.03$ and $a=-0.03$) is specified in plots.
Solid black lines present analytical dependencies
[eqs.~(\ref{eq03})--(\ref{eq05})] with $\la{f}^2\ra\approx0.5464$
and $\la(f')^2\ra\approx0.5775$.  The ratio between the Lyapunov
exponent and diffusion constant is constant and obeys
eq.~(\ref{eq06}) up to the calculation accuracy (see the lower row
of plots).
 }
  \label{fig1}
\end{figure}

In nature and technology, oscillations of systems meant to be
periodic are never perfectly periodic; noise is always present
even for clocks, lasers, electronic generators, etc. On the other
hand, for the majority of systems of practical interest their
natural dynamics is non-overwhelmed by noise. We typically deal
with natural system dynamics affected by noise rather than with
strong noise signal transformed by the system. Therefore, the case
of weak noise can be treated as a typical one, and we restrict our
consideration to this case. For weak noise and feedback the
dynamics can be described within the framework of the phase
description up to the leading order of
accuracy~\cite{Kuramoto-2003,Goldobin_etal-2010}:
\begin{eqnarray}
\dot{\ph}=\Omega_0+a\int\limits_0^{+\infty}
 \sum_{i=1}^{N}\sum_{j=1}^{N}G_{ij}(t_1)H_{ij}\big(\ph(t-t_1),\ph(t)\big)\,dt_1
\nonumber\\
 +\e f\big(\ph(t)\big)\circ\xi(t),\qquad
 \label{eq02}
\end{eqnarray}
where $\Omega_0$ is the natural frequency of the oscillator, $f$
is a $2\pi$-periodic function featuring the sensitivity of the
phase to noise, $\e$ is the noise amplitude, $H_{ij}(\psi,\ph)$ is
the increase of the phase growth rate created by the feedback term
$x_j(\psi)$ acting on the variable $x_i(\ph)$. One can calculate
 $H_{ij}(\psi,\ph)=(\partial\ph/\partial x_i)x_j(\psi)$
for a known phase field $\ph({\bf x})$
(see~\cite{Guckenheimer-1975} and, e.g., \cite{Arai-Nakao-2008}
for examples of calculation of $\ph({\bf x})$; the Maple-program
in supplementary material for~\cite{Goldobin-2011} calculates
$H_{ij}(\psi,\ph)$). We would like to notice that
eq.\,(\ref{eq02}) is not an approximation but an accurate
mathematical description for the case we consider.

The phase of the noise-free oscillator grows uniformly and its
shifts neither decay nor grow in time. Noise creates irregularity
of the phase growth rate, measured by the phase diffusion constant
$D$ (DC): $\lla(\ph(t)-\la\ph(t)\ra)^2\rra\propto D\,t$.
Additionally, noise results in convergence of trajectories, phase
shifts decay, and the exponential rate of this decay is measured
by the Lyapunov exponent $\lambda$ (LE).

In~\cite{Tukhlina_etal-2008} the analytical calculation of DC was
extended to the general case of linear feedback, which covers the
particular cases of simple and recursive delay
feedback~\cite{Goldobin-Rosenblum-Pikovsky-2003,Pawlik-Pikovsky-2006,Goldobin-2011}
and was employed for the case of linear frequency filter. The
derivation procedure for LE~\cite{Goldobin-2008} is essentially
more laborious than that for DC even for the case of simple delay
feedback and is strictly limited to the case of weak noise
(meanwhile the analytical evaluation of DC does not necessarily
require noise to be weak~\cite{Goldobin-Rosenblum-Pikovsky-2003}).
Here we skip the derivation procedure due to its laboriousness and
the fact that it relies on ideas previously
elaborated~\cite{Goldobin-2008}, and provide the results.

For the general case of linear feedback, mean frequency $\Omega$,
phase diffusion constant $D$, and Lyapunov exponent $\lambda$ read
\begin{eqnarray}
 &&
\Omega=\Omega_0+a\int_0^{+\infty}
  \sum_{i=1}^{N}\sum_{j=1}^{N}G_{ij}(t)h_{ij}(-\Omega t)\,dt,
\label{eq03}\\[5pt]
 &&
D=\frac{2\e^2\la{f^2}\ra_\ph}{\Big(1+a\int\limits_0^{+\infty}
 t\sum\limits_{i=1}^{N}\sum\limits_{j=1}^{N}G_{ij}(t)h'_{ij}(-\Omega t)\,dt\Big)^2}
\nonumber\\
 &&
\qquad\qquad\qquad\qquad=2\e^2\la{f^2}\ra_\ph\left(\frac{\partial\Omega}{\partial\Omega_0}\right)^2,
\label{eq04}\\[5pt]
 &&
\lambda=-\frac{\e^2\la(f')^2\ra_\ph}{\Big(1+a\int\limits_0^{+\infty}
 t\sum\limits_{i=1}^{N}\sum\limits_{j=1}^{N}G_{ij}(t)h'_{ij}(-\Omega t)\,dt\Big)^2}
\nonumber\\
 &&
\qquad\qquad\qquad\qquad=-\e^2\la(f')^2\ra_\ph\left(\frac{\partial\Omega}{\partial\Omega_0}\right)^2,
\label{eq05}
\end{eqnarray}
where $\la...\ra_\ph\equiv(2\pi)^{-1}\int_0^{2\pi}...d\ph$,
$h_{ij}(\psi):=\la H_{ij}(\ph+\psi,\ph)\ra_\ph$, and the prime
denotes derivative.  Notice, for any kind of linear feedback the
ratio
\begin{equation}
\frac{-\lambda}{D}=\frac{\la(f')^2\ra}{2\la{f^2}\ra}
\label{eq06}
\end{equation}
is independent of parameters of the feedback.  Previously this
ration was observed only for the case of simple delay feedback.

The modified formulation of the result~(\ref{eq06}) was revealed
also for a certain class of systems where the phase cannot be well
defined. Specifically, in systems below a Hopf bifurcation, noise
can induce oscillatory motion~\cite{Noise-induced-motion-Schoell}
for which the phase is not well defined.
For slightly perturbed periodic oscillations the autocorrelation
function $C_{jj}(s):=\la{x_j(t)x_j(t+s)}\ra$ decays exponentially,
$C_{jj}(s)=c_j(s)\exp(-D|s|)$, where function $c_j(s)$ oscillates
but neither grows nor decays asymptotically.  One can introduce an
alternative quantifier of the irregularity of oscillations, the
correlation time $t_{\rm
corr}:=C_{jj}^{-1}(0)\int_0^\infty|C_{jj}(s)|ds$ and find $t_{\rm
corr}\propto1/D$ for perturbed periodic oscillations.  For
noise-induced oscillations, $t_{\rm corr}$ is an appropriate
quantifier, while $D$ cannot be evaluated. In
Refs.~\cite{Noise-induced-motion-Schoell}, for the van der Pol
oscillator slightly below a Hopf bifurcation, the product $t_{\rm
corr}|\lambda|$ was reported to be independent of the feedback
with time-delay(s), although the feedback significantly changed
$t_{\rm corr}$ and $\lambda$. This is equivalent to the result
(\ref{eq06}).

The validity of our findings (\ref{eq03})--(\ref{eq06}) can be
underpinned with the results of numerical simulation for noisy van
der Pol oscillator
\begin{equation}
\dot{x}=y,\quad
\dot{y}=\mu(1-4x^2)y-x+az(t)+\e\,\xi(t)\,.
\label{eq-vdp}
\end{equation}
Here $\mu$ describes closeness to the Hopf bifurcation point, the
phase $\ph=-\arctan(y/x)$ (for $\mu\ll1$ the limit cycle is:
$x=\cos\ph$, $y=-\sin\ph$). We consider 3 possible cases:
\\
(1) a ``simple'' delay feedback with delay time $\tau$;
\begin{equation}
z^{(1\tau)}(t)=2\big(y(t-\tau)-y(t)\big),
\label{eq-simple}
\end{equation}
which yields
$$
{\bf\hat G}^{(1\tau)}(t)=
\left[\begin{array}{cc}
0&0\\
0&2\big(\delta(t-\tau)-\delta(t-0)\big)
\end{array}\right]
$$
(here we explicitly indicate that
$\int_0^{+\infty}\delta(t-0)dt=1$).
\\
(2) a recursive delay feedback;
\begin{equation}
z^{(\mathrm{m}\tau)}(t)=2\sum_{n=0}^{+\infty}R^n\big(y(t-(n+1)\tau)-y(t-n\tau)\big),
\label{eq-multiple}
\end{equation}
where $|R|<1$, which yields
$$\begin{array}{l}
{\bf\hat G}^{(\mathrm{m}\tau)}(t)\\
\quad=\left[\begin{array}{cc}
0&0\\
0&2\!\sum\limits_{n=0}^{+\infty}\!R^n\big(\delta(t-(n+1)\tau)-\delta(t-n\tau-0)\big)
\end{array}\right].\end{array}
$$
\\
(3) a linear frequency filter;
\begin{equation}
 z^{(\mathrm{lff})}(t)=2\dot{u}(t),\qquad
 \ddot{u}+\alpha\dot{u}+\tilde{\omega}_0^2u=\alpha x(t),
\label{eq-filter}
\end{equation}
which yields
$$
{\bf\hat G}^{(\mathrm{lff})}(t)=
\left[\begin{array}{cc}
0&0\\
2\alpha e^{-\alpha t/2}
 (\cos{\omega_0t}-\frac{\alpha}{2\omega_0}\sin{\omega_0t})&0
\end{array}\right],
$$
$\omega_0=\sqrt{\tilde{\omega}_0^2-(\alpha/2)^2}$.

For $\mu\ll1$, the van der Pol oscillator possesses a circular
limit-cycle and $h_{21}(\ph)=-(1/2)\cos\ph$,
$h_{22}(\ph)=(1/2)\sin\ph$, $f(\ph)=-\sin\ph$. The analytical
results derived with the above formulae match well the results of
numerical
simulation~\cite{Goldobin-Rosenblum-Pikovsky-2003,Pawlik-Pikovsky-2006,Tukhlina_etal-2008},
as can be well seen in Fig.~\ref{fig1}. Remarkably, the ratio
between LE and DC is constant with a good accuracy even when they
vary by one order of magnitude, in agreement with
eq.~(\ref{eq06}).

\begin{figure}[!t]
\center{\sf
(a)\hspace{-4mm}\resizebox{0.33\columnwidth}{!}{%
\includegraphics
 {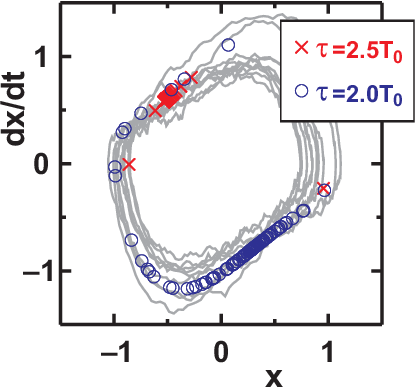} }
 \qquad\qquad
(b)\hspace{-4mm}\resizebox{0.33\columnwidth}{!}{%
\includegraphics
 {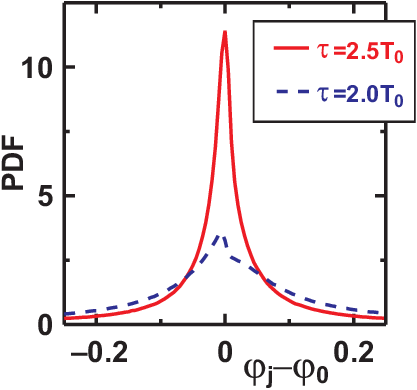} }
\\[15pt]
(c)\hspace{-4mm}\resizebox{0.40\columnwidth}{!}{%
\includegraphics
 {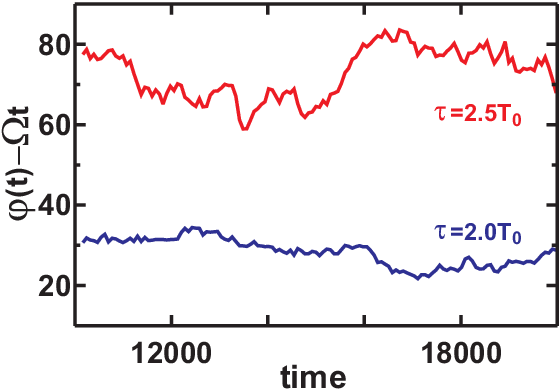} }
}
  \caption{(Color online)
Ensemble of 100 van der Pol oscillators (\ref{eq-vdp}) with
nonidentical frequencies subject to the simple delay feedback
(\ref{eq-simple}). The distribution of frequencies is the gaussian
one centered at $1$ with standard deviation $0.001$, feedback
strength $a=0.03$, noise amplitude $\e=0.1$. (a) The gray line
plots trajectory of one oscillator, the symbols plot snapshots of
the ensemble for two specified values of the delay time (check
these values in Fig.~\ref{fig1}a). (b) The distribution of the
phase deviations from the value $\varphi_0$ corresponding to the
instantaneous ensemble-mean state. (c) The deviation of the phase
from its mean growth for an oscillator (the offset of the vertical
axis is arbitrary).}
  \label{fig2}
\end{figure}

\section{Uncertainty principle}
Constancy of the ratio (\ref{eq06}),
$$
\frac{-\lambda}{D}=\frac{\la(f')^2\ra}{2\la{f^2}\ra},
$$
can be formulated as a kind of uncertainty principle for the
linear feedback control techniques:
\\
\indent{\em The reliability of a noisy oscillator can be
significantly enhanced
 (by means of a weak linear feedback)
but at the price of the loss of its coherence and, vice versa, the
coherence can be significantly enhanced but with the loss of
reliability.}

\begin{figure}[!t]
\center{\sf
(a)\hspace{-4mm}\resizebox{0.33\columnwidth}{!}{%
\includegraphics
 {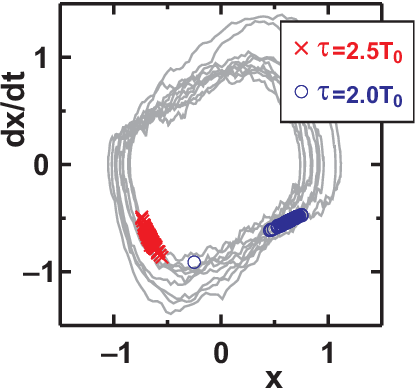} }
 \qquad\qquad
(b)\hspace{-4mm}\resizebox{0.33\columnwidth}{!}{%
\includegraphics
 {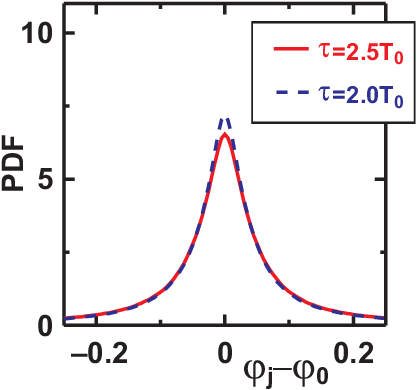} }
\\[15pt]
(c)\hspace{-4mm}\resizebox{0.40\columnwidth}{!}{%
\includegraphics
 {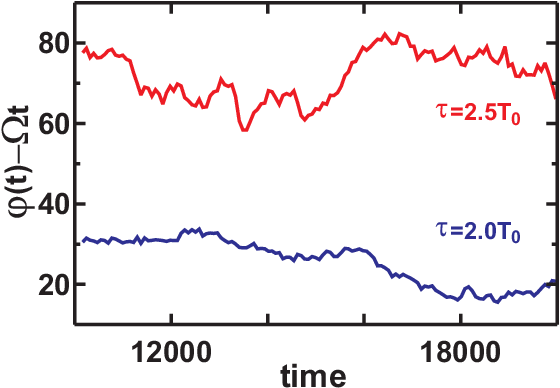} }
}
  \caption{(Color online)
Ensemble of 100 van der Pol oscillators (\ref{eq-vdp}) subject to
the simple delay feedback (\ref{eq-simple}), common noise $\e=0.1$
and intrinsic noise $\e_\mathrm{int}=0.005$. For details see
caption to Fig.~\ref{fig2}.}
  \label{fig3}
\end{figure}

\section{Implications for ensembles of oscillators}
For ensembles of uncoupled oscillators driven by common noise we
know, that the imperfectness of identity of oscillators or
intrinsic noise leads to imperfectness of synchronization. The
characteristic spreading of states
$\propto1/\sqrt{-\lambda}$~\cite{Goldobin-Pikovsky-2005b}. That
is, higher reliability leads to stronger synchrony of
oscillations. Thinking of employment of control techniques we
should distinguish two ``pure'' situations: (i)~slightly
non-identical oscillators driven by identical noise and
(ii)~identical oscillators with small intrinsic noise, individual
for each oscillator.

For an ensemble of non-identical oscillators, the dispersing of
phases is owed to the mismatch of natural frequencies. This
mismatch is nearly unaffected by small feedback, while a
synchronizing action measured by LE is influenced by the feedback.
Hence, the dispersion of states is perceptive to the feedback
control. For instance, one can see in Fig.~\ref{fig2} that the
ensemble is well synchronized (Fig.~\ref{fig2}a,b) when the phase
diffusion is large (Fig.~\ref{fig2}c), and, on the contrary, for a
small diffusion (enhanced coherence) we observe a much poorer
synchrony.

For an ensemble of identical oscillators with intrinsic noise, the
situation is different. The dispersion is owed to the mutual
diffusion of oscillator phases created by intrinsic noise.
Similarly to~\cite{Goldobin-Pikovsky-2005b}, we can find
$\Delta\ph\propto\sqrt{D_\mathrm{int}}/\sqrt{-\lambda}$, where
$D_\mathrm{int}$ characterizes the diffusion due to intrinsic
noise. The diffusion owing to intrinsic noise is expected to be
subjected to the effect of the feedback in the same manner as the
total diffusion. Hence, we expect the feedback to not influence
the phase dispersion,
$\Delta\ph\propto\sqrt{D_\mathrm{int}}/\sqrt{-\lambda}\approx
const$. Indeed, in Fig.~\ref{fig3} one can see that the
distribution of phase deviations is tolerant to feedback control.
In this case, one can employ the feedback control techniques to
control the coherence without the loss of synchrony.

\begin{figure}[!t]
\center{\sf
(a)\hspace{-4mm}\resizebox{0.33\columnwidth}{!}{%
\includegraphics
 {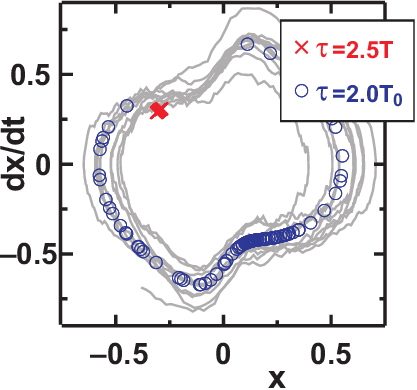} }
 \qquad\qquad
(b)\hspace{-4mm}\resizebox{0.33\columnwidth}{!}{%
\includegraphics
 {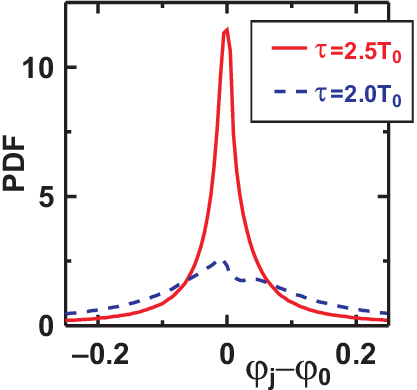} }
\\[15pt]
(c)\hspace{-4mm}\resizebox{0.40\columnwidth}{!}{%
\includegraphics
 {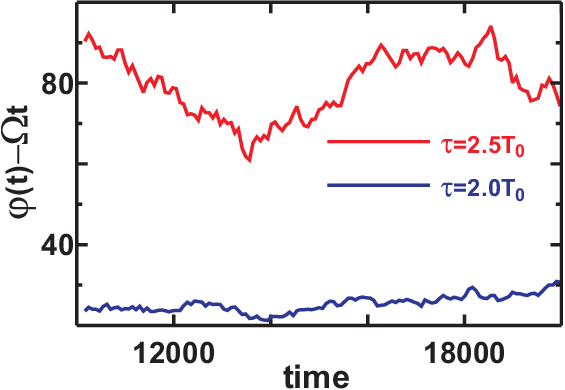} }
}
  \caption{(Color online)
Ensemble of 100 pendulum clock oscillators (\ref{eq-pcl}) with
$\mu=0.4$ and nonidentical frequencies subject to the simple delay
feedback (\ref{eq-simple}). The distribution of
$\widetilde{\Omega}_0$ is the gaussian one centered at $1$ with
standard deviation $0.001$, feedback strength $a=0.03$, noise
amplitude $\e=0.05$. For details see caption to Fig.~\ref{fig2}.}
  \label{fig4}
\end{figure}

\begin{figure}[!t]
\center{\sf
(a)\hspace{-4mm}\resizebox{0.33\columnwidth}{!}{%
\includegraphics
 {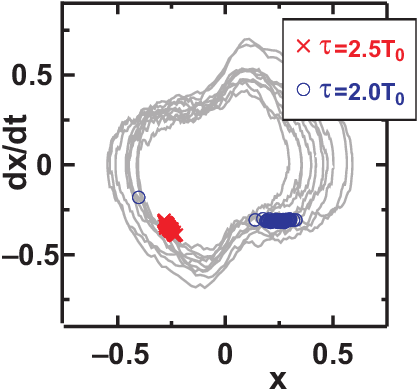} }
 \qquad\qquad
(b)\hspace{-4mm}\resizebox{0.33\columnwidth}{!}{%
\includegraphics
 {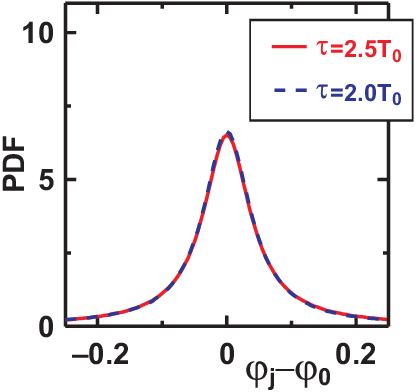} }
\\[15pt]
(c)\hspace{-4mm}\resizebox{0.40\columnwidth}{!}{%
\includegraphics
 {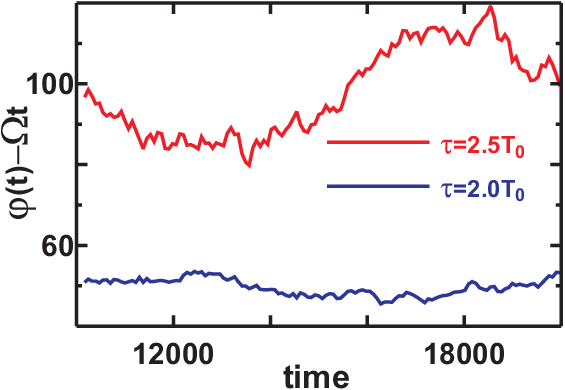} }
}
  \caption{(Color online)
Ensemble of 100 pendulum clock oscillators (\ref{eq-pcl}) with
$\mu=0.4$ subject to the simple delay feedback (\ref{eq-simple}),
common noise $\e=0.05$ and intrinsic noise
$\e_\mathrm{int}=0.0025$. For details see caption to
Fig.~\ref{fig2}.}
  \label{fig5}
\end{figure}

Weakness of noise---which is the condition of accuracy of relation
(\ref{eq06})---is the feature of clock systems. Simultaneously,
the accuracy and synchronization effects are especially important
for clock oscillators. Van del Pol oscillator may be treated as a
rough approximation of the pendulum clock
oscillator~\cite{Kapitaniak_etal-2012}. A more appropriate model
for the pendulum clock oscillator (with feedback control and
noise) is
\begin{equation}
\ddot{x}+\mu\dot{x}+\widetilde{\Omega}_0^2\sin{x}
 =\mathrm{sign}(\dot{x})M(x)+az(t)+\e\,\xi(t)\,,
\label{eq-pcl}
\end{equation}
where function $M(x)$ is localized near
$x=0$~\cite{Kapitaniak_etal-2012}. For demonstration we adopt
$M(x)=M_0x_M^{-1}\exp(-x^2/x_M^2)$ with $x_M=0.1$ and $M_0=0.1$.
In Fig.\,\ref{fig4}, one can see that for the ensemble of
nonidentical systems subject to common noise the synchrony is
strongly influenced by the feedback control as well as the
coherence, and they cannot be enhanced simultaneously. In
Fig.\,\ref{fig5}, for identical oscillators with additional
inherent noise synchrony is unaffected by feedback control, while
the coherence can be significantly enhanced, as well as for Van
der Pol oscillators.

The tentative hypothesis on the response of ensembles to the
control proposed in the concluding remarks of
Ref.\,\cite{Goldobin-2008} turned out to be valid only for the
case of non-identical oscillators with no intrinsic noise.

\section{Conclusion}
Summarizing, we have discovered the universal {\em uncertainty
principle}, which is valid for general noisy limit-cycle
oscillators subject to a general linear feedback control and
proved it both analytically and numerically.  Mathematically, this
uncertainty principle takes the form of eq.~(\ref{eq06}); the
ratio of the Lyapunov exponent ($\lambda$), measuring reliability
({\it or} the response stability,
\cite{Mainen-Sejnowski-1995,Uchida-Mcallister-Roy-2004}),
and the diffusion constant ($D$), measuring coherence ({\it or}
the constancy of the instantaneous frequency in time), is
independent of the noise strength and feedback parameters.  That
is the reliability of a noisy oscillator can be significantly
enhanced by means of a relatively weak linear feedback, but at the
price of the loss of its coherence, and vice versa, the coherence
can be significantly enhanced but with the loss of reliability.
The principle has an implication to practical issues of the
control of synchronization in non-ideal ensembles of oscillators
in nature and
technology~\cite{bacterial_colonies,population_synchrony,Tass-1999,Pikovsky-Rosenblum-Kurths-2001-2003,uncoupled_ensembles-synchronization,globally_coupled_ensembles-control}.
For ensembles of weakly non-identical oscillators driven by common
noise (cf Fig.~\ref{fig2}) the enhancement of synchrony, achieved
due to the reliability increase, results in a poorer coherence of
each individual oscillator. Unsimilarly, for ensembles of
identical oscillators with intrinsic noise (cf Fig.~\ref{fig3})
the synchrony is not influenced by the change of
reliability/coherence; therefore, coherence can be controlled
without stray effects on the synchronization.

Notice, that ensembles of globally coupled oscillators in an
asynchronous state immediately correspond to the case of uncoupled
oscillators receiving common driving; this driving is the
ensemble-mean value forcing an oscillator via the coupling
term~\cite{Topaj-Kye-Pikovsky-2001}. As long as collective modes
vanish, the ensemble-mean value fluctuating about zero may be well
considered as a stochastic process and one can speak of
synchronization by common noise. (When the collective mode
appears, it can be regular in time; synchronization by a regular
signal drastically differs from that by a nonperiodic one,
e.g.~\cite{Ritt-2003,Goldobin-2005}.) Therefore, our findings are
related to the fundamentals of synchronization in ensembles of
globally coupled oscillators as
well~\cite{globally_coupled_ensembles-control,Hramov-Koronovskii-Moskalenko-2005-2006}.


\end{document}